# High-Bandwidth 940 nm VCSEL with Zn-diffusion for Optical Communications

**Fu-He Hsiao** [1,2], **Yu-Jie Lin** [3], **Chia-Jung Tsai** [2,3], **Chia-Chen Li** [2], **Yun-Han Chang** [2], **Chih-Ting Chang** [2], **Jr-Hau He** [4], **Chun-Liang Lin** [1], **Yu-Heng Hong** [2,*], **and Hao-Chung Kuo** [2,3,*]

[1] Department of Electrophysics, College of Science, National Yang Ming Chiao Tung University, Hsinchu, 30010, Taiwan
[2] Semiconductor Research Center, Hon Hai Research Institute, Taipei, 11492, Taiwan
[3] Department of Photonics and Institute of Electro-Optical Engineering, College of Electrical and Computer Engineering, National Yang Ming Chiao Tung University, Hsinchu, 30010, Taiwan
[4] International College of Semiconductor Technology, National Yang Ming Chiao Tung University, Hsinchu, 30010, Taiwan

* Correspondence: Yu-Heng Hong: enoch.yh.hong@foxconn.com, Hao-Chung Kuo: hckuo0206@nycu.edu.tw.

**Abstract**

We present a systematic design methodology, combining simulation and experimental validation, for high-speed 940 nm vertical-cavity surface-emitting lasers (VCSELs). A comprehensive simulation study was conducted to optimize the device structure, focusing on the number of oxide layers and the aperture size, which predicted a maximum modulation bandwidth of over 35 GHz. To validate this approach, an optimized device with a 4-μm double-oxide aperture was fabricated and characterized. Crucially, during the fabrication process, a Zn-diffused region was incorporated to further enhance device performance. The experimental results demonstrate a modulation bandwidth of 34 GHz and successful 100 Gbit/s PAM-4 data transmission. The excellent agreement between the simulated and measured performance validates the effectiveness of our design methodology, providing a reliable framework for developing next-generation optical interconnects.

## 1. Introduction

The demand for data transmission bandwidth is growing rapidly, driven by hyperscale data centers, 5G/6G deployment, and data-intensive applications. Artificial intelligence and machine learning workloads pose a particular challenge, requiring massive parallel processing with low-latency, high-throughput interconnects [1,2]. As traditional copper links fall short, energy-efficient optical interconnects become essential. This underscores the need for advanced components like the vertical-cavity surface-emitting lasers (VCSELs) studied here, capable of supporting high-speed communication and preventing bottlenecks in next-generation computing and networks[3,4].

VCSELs have emerged as one of the most versatile semiconductor laser technologies due to their unique structural and performance advantages over conventional edge-emitting lasers (EELs). Unlike EELs, which emit light from the cleaved chip edge, VCSELs emit perpendicular to the wafer surface, enabling on-wafer testing, high fabrication yield, and straightforward integration into two-dimensional arrays [5]. In addition, VCSELs typically exhibit low threshold currents, circular beam profiles, and excellent coupling efficiency into optical fibers or free space. These characteristics have established VCSELs as compact, energy-efficient, and cost-effective light sources for both data communication and sensing applications [6-8]. Furthermore, enhancing the power conversion efficiency of VCSELs remains a key area of ongoing research, aiming to further reduce the power consumption of optical systems [9].

Among the available wavelengths, 940 nm VCSELs have gained particular importance for applications such as 3D sensing, eye-tracking, biometric recognition, and automotive LiDAR [10-13]. The 940 nm wavelength lies in the near-infrared region, where silicon-based detectors provide high responsivity while minimizing interference from visible ambient light [14]. This not only improves sensing accuracy in complex environments but also ensures compatibility with CMOS technology, accelerating their adoption in consumer electronics and automotive systems. When compared to the more mature 850 nm VCSELs, which are widely deployed in short-reach optical interconnects [15-17], 940 nm devices face unique material and design challenges. Specifically, achieving 940 nm emission requires indium incorporation into the multiple quantum wells (MQWs), introducing additional strain and growth complexity. Nevertheless, this adjustment enables precise bandgap tuning and provides key advantages: higher indium composition enhances carrier confinement and differential gain [18-21], which directly supports extended modulation bandwidth and high-speed operation. Moreover, indium incorporation has been shown to suppress the formation of dark-line defects (DLDs), one of the major reliability concerns in III–V lasers, thereby significantly improving device stability and prolonging operational lifetime [22,23]. Thus, while 940 nm VCSELs present greater epitaxial challenges compared to 850 nm devices, they offer significant opportunities in next-generation applications that demand both high-speed performance and long-term reliability. Indeed, advancing VCSEL technology to meet the demands of future data networks remains a key research focus, with recent efforts targeting both the enhancement of device characteristics and the improvement of energy efficiency at high speeds [22,24,25].

While record-breaking bandwidths have been reported in the literature, the primary goal of this work is not to set a new performance benchmark, but rather to present a validated design methodology that connects structural parameters to high-speed performance. Therefore, we systematically investigate the impact of oxide aperture size and the number of oxide layers on the modulation performance of 940 nm VCSELs. Through detailed design and optimization, we demonstrate devices capable of achieving a 3-dB small-signal modulation bandwidth exceeding 27 GHz across a wide temperature range. To further bridge the gap between simulation and high-speed realization, a Zn-diffused region was strategically implemented during fabrication to minimize p-DBR resistance and RC parasitics. Consequently, experimental measurements confirm a bandwidth of 34 GHz and successful PAM-4 data transmission at 100 Gb/s per lane. These results highlight the potential of 940 nm VCSELs not only for high-speed optical interconnects but also for emerging consumer and automotive applications that require compact form factors, energy efficiency, and robust performance under demanding operating conditions.

## 2. Device Structure Design

In this work, the 940 nm VCSEL is simulated using Crosslight PICS3D software to analyze its structural design and performance. As shown in Figure 1, the epitaxial structure is grown on a GaAs substrate and features a short cavity design with an optical length of $\lambda/2n$, which is crucial for achieving high optical confinement and realizing high-speed operation. This optical resonator is formed by top and bottom distributed Bragg reflectors (DBRs). The bottom n-DBR consists of 34 pairs of n-doped $Al_{0.08}Ga_{0.92}As/Al_{0.9}Ga_{0.1}As$, while the top p-DBR is composed of 17 pairs of the same materials with p-type doping. With each layer's thickness, $d$, following the quarter-wavelength design ($d=\lambda/4n$), the DBRs provide the high reflectance required for optical feedback and laser oscillation, which is fundamental to the device's vertical emission and suitability for array fabrication. Positioned at the center of this resonant cavity is the active region, composed of three pairs of strained multiple quantum wells (MQWs). This MQW structure consists of 5.6 nm-thick $In_{0.195}Ga_{0.805}As$ quantum wells separated by 10 nm-thick $Al_{0.2}Ga_{0.8}As$ barriers. The implementation of a strained MQW design is advantageous as it provides high differential gain, low dislocation density, and a broad modulation bandwidth. To further enhance device performance, an oxide confinement layer is incorporated within the top p-DBR. This is achieved by selectively oxidizing a specific AlGaAs layer with a higher Al composition in a high-temperature steam environment, thereby forming an oxide aperture, and providing outstanding lateral electrical and optical confinement, effectively funneling carriers into the active region to reduce threshold current. While the initial device parameters were optimized through the aforementioned simulation framework, a Zn-diffused region was strategically implemented during the fabrication stage to further enhance experimental performance.[29] This process was specifically employed to significantly reduce the resistance of the top p-DBR layers, thereby decreasing the RC time constant for high-speed operation while simultaneously suppressing the transverse modes of the VCSEL to improve overall signal integrity.

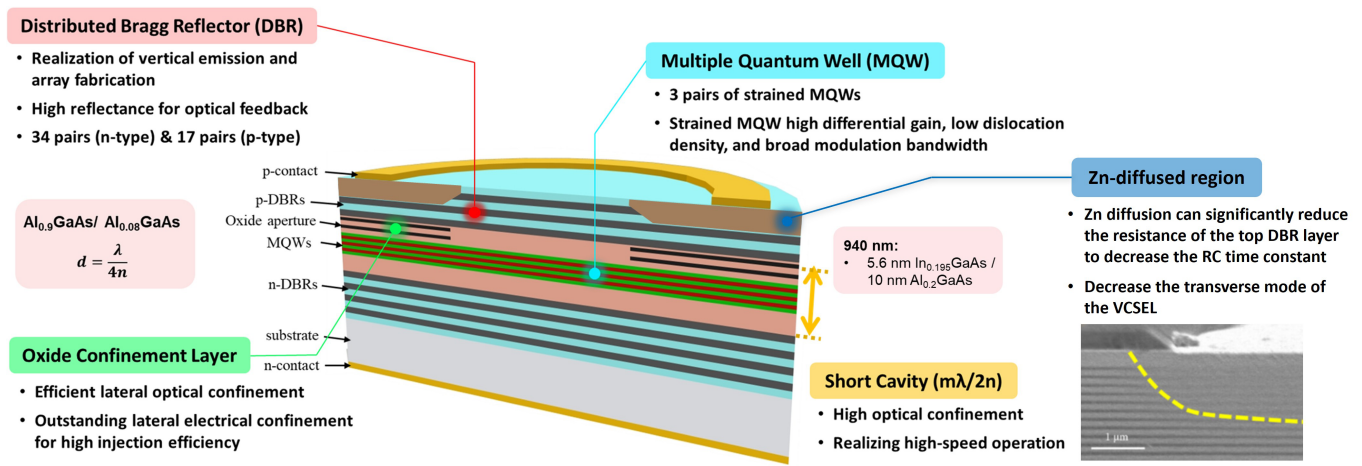

**Figure 1.** Schematic illustration of 940 nm VCSEL structure.

## 3. Results and Discussion

The modulation bandwidth of a VCSEL is co-limited by its intrinsic carrier dynamics and extrinsic electrical parasitics. These parasitics created an RC time constant that fundamentally restricts the maximum speed. Therefore, a primary goal of our design optimization is to mitigate these parasitic effects. In the following sections, we investigate how the number of oxide layers and the aperture size can be engineered to reduce the RC limitation and maximize the modulation bandwidth.

### 3.1. Number of Oxidation Layers

At the first step in our design optimization, we investigate the impact of the number of oxide confinement layers on the device's high-speed performance by simulating and comparing the modulation bandwidth under different structures that incorporate single-, double-, and triple-layer oxide configurations. Figures 2(a) shows the schematic diagram of the oxidation layer function and the left-hand side of Figures 2(b)-(d) illustrate the internal standing wave optical field distribution for these three structures, indicating the position of the oxide layers (gray regions) relative to the standing wave. As clearly shown, a key design principle is the strategic placement of the oxide layers precisely at the nodes of the standing wave's electric field. Since the electric field intensity is minimal at these nodal positions, the oxide layers cause only a slight disturbance to the optical mode profile, leaving the macroscopic standing wave pattern essentially unaffected across the various configurations. This positioning is critical as it minimizes optical scattering and absorption losses, thereby preserving the high quality factor of the resonant cavity [26,27].

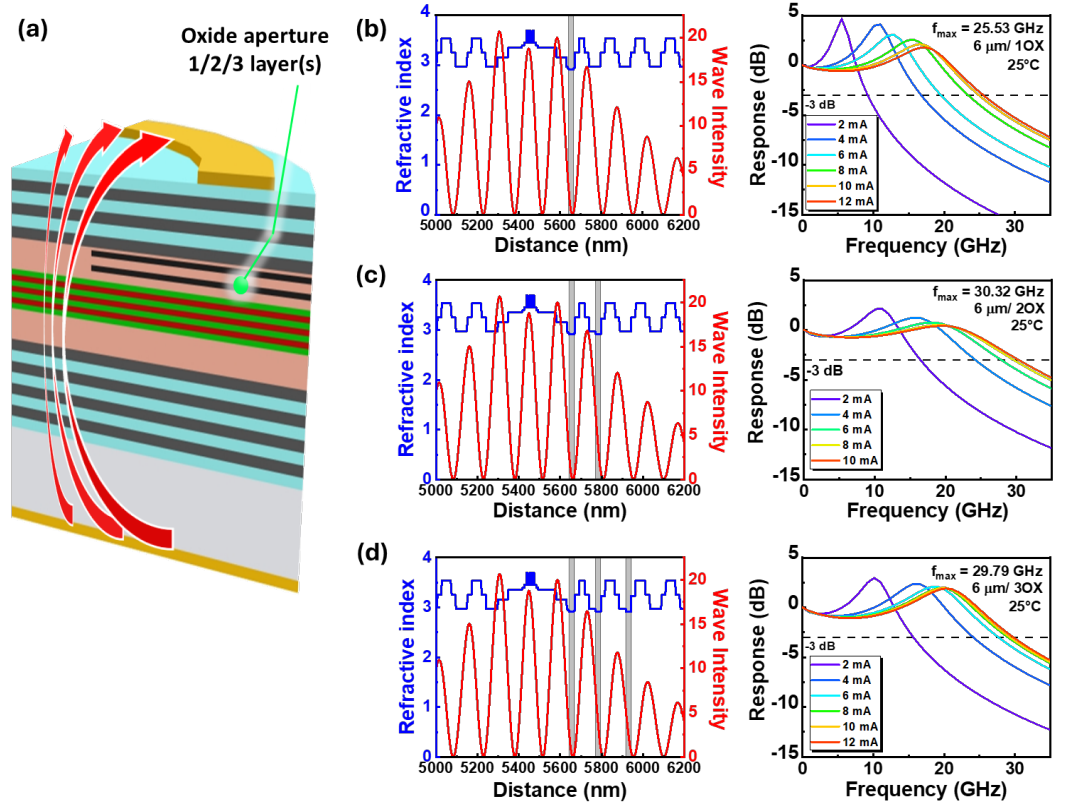

**Figure 2.** (**a**) The schematic diagram of the oxidation layer function. The optical field distribution and frequency response of the VCSEL with (**b**) one-layer, (**c**) two-layer, and (**d**) three-layer oxidation.

The motivation for exploring multi-oxide-layer designs is to reduce the parasitic capacitance that limits high-speed performance. While a single oxide layer already effectively defines the current injection area, additional oxide layers can enhance the lateral confinement of the electrical current. This stronger confinement effectively reduces the active area of the p-n junction, thereby lowering the total device capacitance. However, this approach involves a critical trade-off, as each additional layer tends to increase the total series resistance and introduce more optical scattering loss from the additional aperture, degrading the cavity's Q-factor [26], while the benefits of capacitance reduction diminish. An optimal design must therefore balance these competing effects to minimize the overall RC time constant. The right-hand side of Figures 2(b)-(d) present the small-signal modulation response results that directly illustrate this trade-off. While the modulation bandwidth increases with the injection current in all configurations, a direct comparison reveals that the double-layer oxide configuration consistently exhibits the best frequency response, achieving the highest modulation bandwidth across all tested current levels. This result indicates that while the triple-layer structure may offer slightly lower capacitance, its performance is ultimately hindered by the aforementioned detrimental factors of increased resistance and optical loss, which outweigh any marginal capacitance benefit. Therefore, these simulation results confirm that a double-layer oxide confinement structure represents an optimized design for maximizing the high-speed modulation capabilities of this particular VCSEL, a finding that is consistent with design trends observed in high-speed VCSELs [28-32].

*3.2. Oxide Aperture Size*

Following the optimization of the design to a double-layer oxide structure, we conducted a detailed investigation into the influence of the oxide aperture size on the VCSEL's overall performance. This analysis covers devices with apertures of 10, 8, 6, and 4 μm. The most significant impact was observed in the high-speed modulation performance. As illustrated in Figure 3, reducing the aperture size yields a dramatic enhancement in modulation speed. A clear trend was observed where smaller apertures reached a higher maximum bandwidth at a lower bias current. The maximum -3 dB bandwidth for the 10 μm device was 23.41 GHz, achieved at a bias of 18 mA, while the 4 μm device achieved a significantly higher bandwidth of 35.68 GHz at lower bias of 8 mA. This improvement is attributed to two primary mechanisms. First, a smaller aperture provides stronger current and optical confinement, which leads to a higher carrier density in the active region and an enhanced differential gain. Second, and critically for high-frequency operation, the reduced aperture size significantly lowers the device's parasitic capacitance. Shrinking the aperture reduces the effective area for charge accumulation, decreasing the total capacitance. This directly lowers the fundamental RC time constant, which enables a faster electrical response and a higher modulation bandwidth.

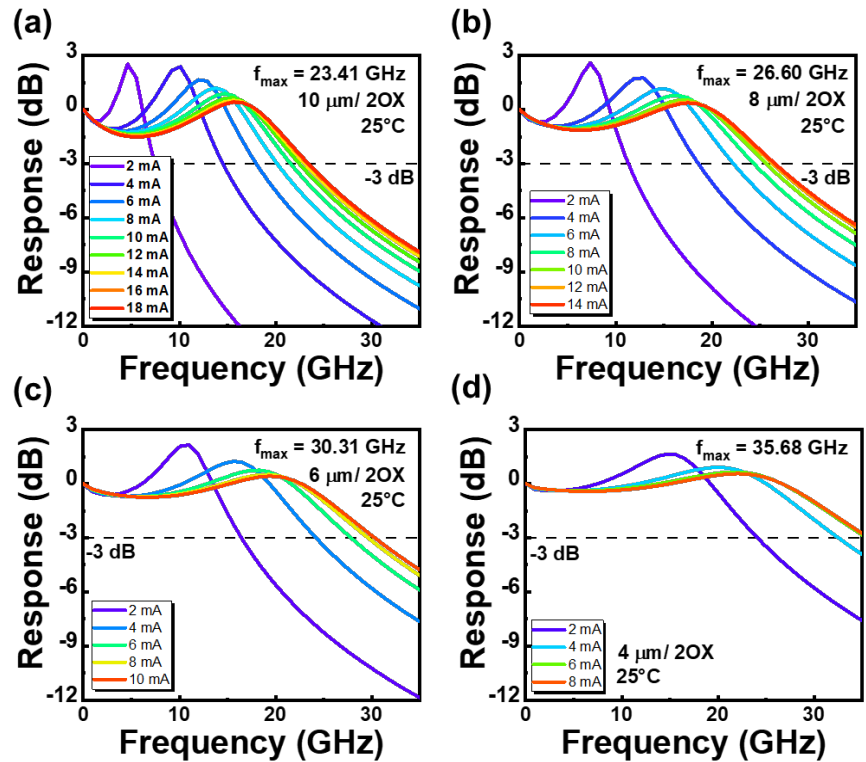

**Figure 3.** Frequency response of 940 nm VCSEL under different aperture sizes. (**a**) 10 μm, (**b**) 8 μm, (**c**) 6 μm, and (**d**) 4 μm.

While beneficial for speed, shrinking the aperture also affects the static characteristics. As shown in Figures 4 (a) and (b), the simulated Power-Current (P-I) curves show that devices with smaller apertures yield a lower threshold current due to the improved carrier confinement. However, this tight confinement comes at a cost. The devices with smaller aperture sizes also exhibit an earlier onset of thermal rollover, a consequence of higher current density and more severe self-heating effects. This is consistent with the differential resistance, which increases as the aperture shrinks.

The high-speed performance of the VCSELs under varying aperture sizes can be further characterized by their modulation current efficiency factor (MCEF), often referred to as the D-factor. This crucial metric quantifies how efficiently the injection current is converted into modulation bandwidth, and it is defined by the equation [33]:

$$MCEF = \frac{f_{-3\,dB}}{\sqrt{I-I_{th}}}, \qquad (1)$$

where $f_{-3\,dB}$ represents the -3 dB bandwidth and $I_{th}$ represents the threshold current. As illustrated in Figure 4 (c), the MCEF is directly proportional to the slope of the linear fit for each aperture size. It is immediately evident that the 4 μm aperture device exhibits the steepest slope, indicating the highest MCEF among all simulated designs. A higher MCEF is particularly desirable as it means a target bandwidth can be achieved with less drive current, leading to lower power consumption. This strong correlation underscores that a smaller oxide aperture not only achieves the highest absolute -3 dB bandwidth but also translates to superior current efficiency in driving the high-speed modulation, making the 4 μm design the most efficient for high-frequency operation. In essence, the analysis reveals a classic engineering trade-off, maximizing modulation efficiency versus achieving higher maximum output power and better thermal performance, making the 4 μm design the most efficient choice specifically for high-speed operation.

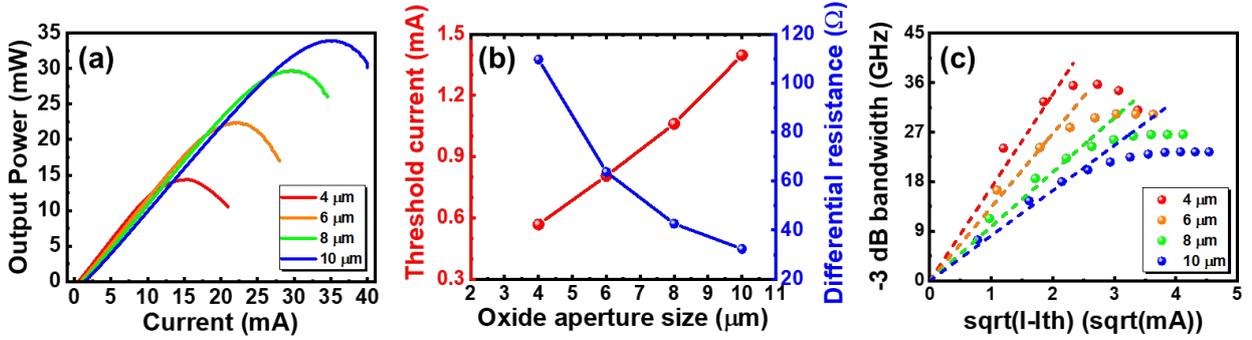

**Figure 4.** Simulation results of (**a**) Power-Current (P-I) characteristics, (**b**) threshold current and differential resistance, and (**c**) -3 dB bandwidth in different oxide aperture sizes.

*3.3. Temperature-Dependent Performance of 4-μm VCSEL Design*

Given its superior frequency response, the 4-μm aperture double-oxide VCSEL was selected for a detailed investigation into its temperature-dependent characteristics. The analysis revealed how elevated temperatures impact the device's static and dynamic performance.

As the ambient temperature increases from 25 °C to 85 °C, several degradation effects are observed. As shown in Figure 5 (a), the P-I characteristics exhibit a higher threshold current and a lower peak output power, which is a result of earlier thermal rollover. This occurs because elevated temperatures reduce the material gain efficiency and exacerbate both gain saturation and thermal carrier recombination, making the laser less efficient at converting electrical current into light. Concurrently, the emission wavelength, shown in Figure 5 (b), exhibits a predictable and clear redshift. Over the operating range from 25 °C to 85 °C, the peak emission wavelength shifts from 942.09 nm to 944.64 nm, corresponding to a thermal tuning coefficient of approximately 0.0425 nm/°C. This is a fundamental semiconductor property attributed to temperature-induced bandgap shrinkage and an increase in the refractive index of the cavity materials.

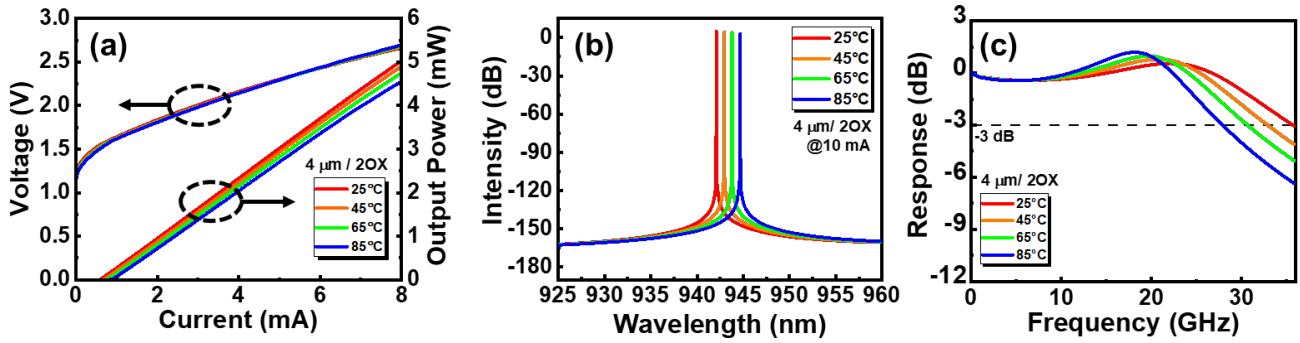

**Figure 5.** Temperature-dependent simulated performance of (**a**) power-current-voltage (P-I-V) characteristics, (**b**) spectra, and (**c**) frequency response of the 4-μm VCSEL.

These underlying physical phenomena also degrade the high-speed modulation capabilities. These underlying physical phenomena also degrade the high-speed modulation capabilities. At elevated temperatures, several thermally-activated mechanisms combine to limit the device's speed. Firstly, increased lattice scattering causes carrier mobility to decrease, which impedes the efficient transport of carriers into the active region. Secondly, non-radiative recombination processes, such as Auger and Shockley-Read-Hall, become more significant. These processes consume injected carriers without generating photons, which not only reduces the overall photon generation efficiency but also directly lowers the differential gain of the quantum wells and can lead to increased carrier leakage. The combination of these factors increases the damping of the system, causing the device to respond more sluggishly to the modulation current. As a result, the relaxation oscillation frequency is suppressed, leading to an inevitable decrease in the maximum modulation bandwidth as temperature rises. Nevertheless, the device demonstrates robust thermal stability. As shown in Figure 5 (c), it is noteworthy that even at a high operating temperature of 85 °C, the VCSEL maintains an impressive modulation bandwidth exceeding 27 GHz. This high level of performance at elevated temperatures confirms its suitability for high-performance applications in demanding thermal environments.

*3.4. 940 nm VCSEL Measurements*

Based on our simulation analysis and optimization results, a 940 nm VCSEL with double-layer 4-μm oxide aperture was fabricated. The 940 nm VCSEL epitaxial structure was first grown on an n-type GaAs substrate using metal-organic chemical vapor deposition (MOCVD), followed by a Zn-diffusion process. The fabrication process began with the P-type ring metal (Ti/Pt/Au), which was deposited using an electron-beam evaporator. After the P-metal deposition, the mesa was etched down to the N-DBR using inductively coupled plasma-reactive ion etching (ICP-RIE), followed by a wet oxidation process in a furnace tube. Subsequently, a ring-shaped N-type metal stack (Au/Ge/Ni/Au) was deposited and annealed to form ohmic contacts. Benzocyclobutene (BCB) was then spin-coated and cured to achieve planarization, after which it was etched back using RIE. Contact holes for both the P and N electrodes were defined by photolithography. Finally, Ti/Au was deposited to form a coplanar RF electrode. Figure 6 (a) shows the top-view optical microscope image of the completed device. It features a ground-signal-ground (GSG) coplanar pad layout, which is specifically designed to minimize parasitic inductance and ensure excellent signal integrity during on-wafer probing. Figure 6 (b) shows the corresponding optical spectrum measured at ambient conditions, exhibiting a peak wavelength of 941.54 nm. Figure 6 (c) illustrates the P-I-V characteristics of the VCSEL at ambient temperature. All measured results are in good agreement with the simulation predictions.

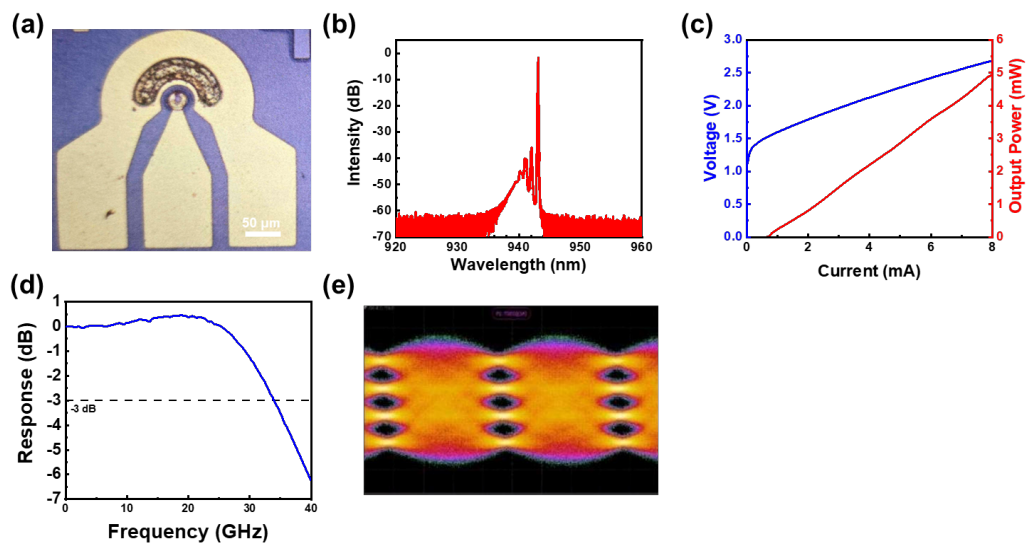

**Figure 6.** (**a**) Top-view microscopic photograph of the VCSEL chip. (**b**) Optical spectrum of the VCSEL at 25 °C. (**c**) Measured P-I-V characteristics of the VCSEL. (**d**) Frequency response of the VCSEL and (**e**) PAM-4 eye diagram at 25 °C and 8 mA bias current.

The device was then characterized at room temperature. All high-speed characterization was performed via on-wafer probing. A high-frequency GSG microwave probe (MPI T67A GSG100) was used to deliver the DC bias and AC modulation signal from a bias-tee to the device pads. The optical output was collected using a customized cone-lensed multi-mode fiber pigtail (~0.5 m length), which was actively aligned to the aperture for maximum coupling efficiency. For small-signal measurements, the optical signal was analyzed by a network analyzer (Agilent N5227B). For the back-to-back (BtB) PAM-4 transmission test, the signal was generated by an arbitrary waveform generator (AWG, Keysight M8194A) and the resulting optical eye diagram was captured by a digital communication analyzer (DCA, Keysight N1046A). As shown by the small-signal modulation response in Figure 6 (d), the VCSEL exhibits an impressive -3 dB bandwidth of 34 GHz. This result is in excellent agreement with our simulation predictions, confirming the accuracy of our physical models and the effectiveness of the optimized design approach. Leveraging this high bandwidth, we conducted a 4-level pulse amplitude modulation (PAM-4) signal transmission test. A 100 Gbit/s PAM-4 signal was applied to the VCSEL biased at 8 mA. The resulting optical eye diagram, captured by an oscilloscope, is depicted in Figure 6 (e). The diagram clearly shows four distinct, well-seperated levels and wide-open eyes, which qualitatively indicate a high signal-to-noise ratio and low signal distortion at this demanding data rate. Quantifying this excellent performance, the measured transmitter and dispersion eye closure of PAM-4 (TDECQ) is 2.1 dB. This successful transmission test verifies the device's capability to achieve data rates exceeding 100 Gbit/s. It is worth noting that these high-speed results were achieved with an aggressive proof-of-concept design, which operated at a high voltage and current density; further optimization would be required to co-optimize for power efficiency and long-term reliability in commercial applications. This demonstrates its significant potential for next-generation high-speed fiber-optic communication applications.

As illustrated in the year-over-year bandwidth benchmark in Figure 7, the primary advantage of this work lies in achieving a superior intrinsic modulation bandwidth of 34 GHz solely through refined structural optimization. While typical 940 nm VCSEL designs leveled off between 22 and 30 GHz [34-40], and latest studies often require complex multi-aperture arrays to push frequency limits [41], our 4-μm double-oxide design maximizes bandwidth by precisely balancing parasitic capacitance and differential gain. This confirms that our methodology provides a more streamlined and efficient path to high-speed performance compared to current high-complexity alternatives.

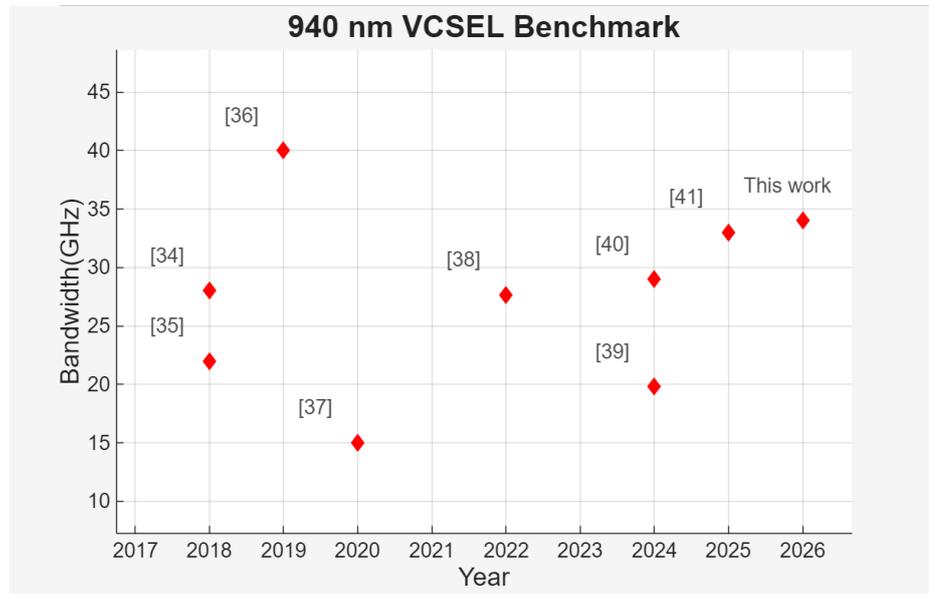

**Figure 7.** Year-over-year bandwidth benchmark of recent 940 nm VCSEL studies.

## 4. Conclusions

In this work, we have presented and experimentally validated a systematic design methodology for 940 nm VCSEL tailored for high-speed applications. Through detailed simulation, we identified an optimized device structure featuring a double-layer 4-μm oxide aperture, which provides an ideal balance between parasitic capacitance reduction and optical loss management. This optimized design was predicted to achieve a maximum modulation bandwidth of over 35 GHz and maintain robust thermal stability with a bandwidth exceeding 27 GHz at temperatures as high as 85 °C. These simulation results were successfully confirmed through the fabrication and characterization of the device. In the experimental phase, a Zn-diffused region was strategically integrated during fabrication to further optimize device performance beyond the initial simulation parameters. The experimental measurements demonstrated a -3 dB bandwidth of 34 GHz at room temperature and successful 100 Gbit/s PAM-4 data transmission. The excellent agreement between our modeling and experimental data validates the effectiveness of our design methodology. The achievement of 34 GHz bandwidth at the 940 nm wavelength not only highlights the potential of these devices for next-generation short-reach optical interconnects but also enhances their utility in advanced applications such as automotive LiDAR and 3D sensing, where high modulation speed can enable superior resolution and faster data acquisition.

**Author Contributions:** Conceptualization, C.-L.L. and H.-C.K.; methodology, Y.-H.H.; software, Y.-J.L.; validation, J.-H.H. and H.-C.K.; formal analysis, F.-H.H., Y.-J.L., C.-J.T., C.-C.L., Y.-H.C., and C.-T.C.; investigation, F.-H.H.; resources, Y.-H.C, Y.-H.H. and H.-C.K.; data curation, F.-H.H. and Y.-J.L.; writing—original draft preparation, F.-H.H.; writing—review and editing, C.-C.L. and Y.-H.H.; visualization, H.-C.K.; supervision, C.-L.L., Y.-H.H. and H.-C.K.; project administration, H.-C.K.; funding acquisition, H.-C.K. All authors have read and agreed to the published version of the manuscript.

**Funding:** This research was funded by the National Science and Technology Council (NSTC) of Taiwan under Grant No. NSTC 113-2622-8-A49-013-SB and No. NSTC 114-2622-8-A49-011-.

**Institutional Review Board Statement:** Not applicable.

I**nformed Consent Statement:** Not applicable.

**Data Availability Statement:** The data presented in this study are available from the corresponding author upon reasonable request.

**Acknowledgments:** The author would like to express profound gratitude to Professor Connie Chang (UC Berkeley), Professor Tien-Chang Lu (NYCU), and Professor Milton Feng (UIUC) for their invaluable guidance, insightful discussions, and continuous support throughout this research. Special thanks are also extended to WinSemi for providing essential technical resources and support, which were instrumental in the completion of this work.

**Conflicts of Interest:** The authors declare no conflicts of interest.